# Optimizing Solar Energy Production in the USA: Time-Series Analysis Using AI for Smart Energy Management

Istiaq Ahmed[1], Md Asif Ul Hoq Khan[2], MD Zahedul Islam[3], Md Sakibul Hasan[4], Tanaya Jakir[5], Arat Hossain[6], Joynal Abed[7], Muhammad Hasanuzzaman[8], Sadia Sharmeen Shatyi[9], Kazi Nehal Hasnain[10]

## Abstract

*As the US rapidly moves towards cleaner energy sources, solar energy is fast becoming the pillar of its renewable energy mix. Even while solar energy is increasingly being used, its variability is a key hindrance to grid stability, storage efficiency, and system stability overall. Solar energy has emerged as one of the fastest-growing renewable energy sources in the United States, adding noticeably to the country's energy mix. Retrospectively, the necessity of inserting the sun's energy into the grid without disrupting reliability and cost efficiencies highlights the necessity of good forecasting software and smart control systems. The dataset utilized for this research project comprised both hourly and daily solar energy production records collected from multiple utility-scale solar farms across diverse U.S. regions, including California, Texas, and Arizona. Training and evaluation of all models were performed with a time-based cross-validation scheme, namely, sliding window validation. Both the Random Forest and the XG-Boost models demonstrated noticeably greater and the same performance across each of the measures considered, with relatively high accuracy. The almost perfect and equal performance by the Random Forest and XG-Boost models also shows both models to have learned the patterns in the data very comprehensively, with high reliability in their predictions. By incorporating AI-powered time-series models like XG-Boost in grid management software, utility companies can dynamically modify storage cycles in real-time as well as dispatch and peak load planning, based on their predictions. AI-powered solar forecasting also has profound implications for renewable energy policy and planning, particularly as U.S. federal and state governments accelerate toward ambitious decarbonization goals. To bridge existing limitations and achieve improved forecast accuracy, future research should investigate hybrid deep models, specifically designs including Long Short-Term Memory (LSTM) and Convolutional Neural Networks with LSTM (CNN-LSTM) architectures. Moreover, future innovation needs to include solar forecasting integration with energy storage modeling and direct real-time deployment in microgrids.*

---

[1] Master of Science in Information Technology, Southern New Hampshire University.
[2] Master of Science in Information Technology, Southern New Hampshire University.
[3] Master of Science in Cybersecurity, Mercy University.
[4] Information Technology Management, St Francis College.
[5] Master's in Business Analytics, Trine University.
[6] Information Technology Management, St Francis College.
[7] Master of Architecture, Miami University, Oxford, Ohio.
[8] Master's in Strategic Communication, Gannon University, Erie, PA, USA
[9] Master of Architecture, Louisiana State University
[10] Master's of Science in Information Technology (MSIT), Westcliff University, Irvine, CA

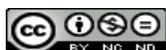

## Introduction

Solar energy has emerged as one of the fastest-growing renewable energy sources in the United States, adding noticeably to the country's energy mix. The U.S. has installed a total of over 160 gigawatts (GW) of photovoltaic (PV) capacity as of 2024, enough to supply about 30 million homes, as reported by the *Solar Energy Industries Association* (SEIA) (Alam et al., 2025; Barua et al., 2025). This expansion has been driven by declining installation costs, reduced by almost 70% in the last decade, coupled with strong federal incentives such as the Investment *Tax Credit (ITC)*. Still, despite this progress, solar energy has a very high variability since it relies on sunlight levels that depend on the time of day, cloud coverage, and atmospheric conditions (Abubakar et al., 2024; Adeoye et al., 2024;) Such variations tend to cause challenges in the reliability of supplying even power output, especially where the penetration of the sun's energy in a region is high as in California, Texas, and Arizona (Chouksey et al., 2025).

The variability of the sun's energy makes it challenging for grid operations personnel to effectively match supply and demand. Historically designed traditional grid systems were not equipped to handle uncertain input in this manner, and as a result, would experience the potential for overgeneration when the sun shines brightly and underproduction when it's cloudy or dark (Cho & Venkateswaran, 2024). Utilities must either resort to expensive Peaker plants, disconnect from existing generators when their output isn't needed, or overinvest in battery capacity, which introduces economic inefficiencies. Additionally, grid congestion happens when excessive photovoltaic-generated energy enters the grid and isn't properly forecasted and regulated (Bouquet et al., 2024).

Retrospectively, the necessity of inserting the sun's energy into the grid without disrupting reliability and cost efficiencies highlights the necessity of good forecasting software and smart control systems. This challenge assumes added urgency in light of the *U.S. Department of Energy's (DOE)* "Solar Futures Study," which estimates that solar might provide as much as 45% of the country's electricity by 2050 (Arevalo & Jurado, 2024). Achieving this goal will not only be an enormous expansion of solar deployment but also the emergence of advanced tools for managing variability. Utilization of AI-enabled time-series analysis for forecasting solar output is thus not merely a technical upgrade—it's a strategic imperative. Firms like *IBM*, through its *IBM Watson AI* platform, and *Google DeepMind*, through renewable energy forecasting tools for wind and solar farms, already illustrate the potential for machine learning-enhanced energy forecasting and grid stability (Chen et al., 2023). These innovations speak to a pivotal moment when data science and clean energy innovation must intersect to create a stronger and less wasteful electricity system.

## Problem Statement

Adeoye et al. (2024), found that the intermittence inherent in the production of solar energy creates uncertainty that can undermine grid stability and minimize the effectiveness of optimizing energy storage deployment. As an example, a precipitous decline in solar irradiance due to rapidly developing cloud coverage can create sudden shifts in energy production and make dynamic energy balancing challenging in real-time. In California, where the country's largest deployment of solar capacity resides, the so-called "duck curve" exemplifies the challenge wherein solar production rises in the middle of the day but peak demand spikes in the evening, requiring substantial ramping of traditional generation resources (Dellosa & Palconit, 2021). The *California Independent System Operator* (CAISO) estimates the misalignment could cost the state billions in inefficiencies unless mitigated through accurate forecasting and advanced grid tools (Gazi et al., 2025).

Furthermore, utility operators today rely on weather forecasts or simple statistical procedures that do not capture the nonlinear and multivariate characteristics of solar energy dynamics. These procedures tend to lose 20–40% of their predictive power under turbulent weather conditions. Without reliable forecasts, utilities will be unable to properly dispatch energy and will experience higher levels of curtailment or wasteful cycling of energy storage systems (Hayajneh et al., 2024). In 2022 itself, California curtailed over 2.3 terawatt-hours (TWh) of renewables, a majority of which were from solar, because they overproduced and were not anticipated. The inefficiency not only wastes clean energy but also challenges investor and consumer confidence in renewables infrastructure (Hasanuzzaman et al., 2025).

Hossain et al. (2024), argued that solving these challenges demands forecasting models that are resilient, adaptive, and can learn from rich sources of data. AI-enabled time-series forecasting fulfills those demands through the provision of highly accurate forecasts and real-time adaptability. Deep neural network models, especially those utilizing LSTM and attention mechanisms, achieved 10-15% reductions in their test deployments as MAPEs. As those models become more advanced and embedded in energy management systems, they provide a means of curtailing, optimizing the deployment of storage resources, and reducing reliance on fossil fuel backup systems. Jakir et al. (2023), posited that companies at the cutting edge of the industry, such as *AES Corporation and Fluence Energy*, already integrate those technologies in their grid-scale battery deployments for better synchronization of their profiles from solar generation resources.

### Relevance in the U.S. Energy Sector

The application of AI-driven time-series forecasting in the context of solar energy management has direct relevance to the development of smart grid infrastructure in the United States (Khan et al., 2023). Smart grids are computerized energy systems utilizing real-time information, communication networks, and automation for dynamically controlling the flow of electricity through production, transmission, and consumption. Smart grid nationwide buildouts could save as much as 9% in energy efficiency by 2030, amounting to billions of dollars in saved energy spending, according to the *U.S. Department of Energy (DOE)*. AI-enhanced solar forecasting forms a central component of this change and provides utilities a means for forecasting the variability of solar output levels, smoothing voltage and frequency levels, and pre-emptive load balancing regimes (Hossain et al., 2025a). Some of the top companies involved in developing AI-integrated grid management solutions are *Schneider Electric, Siemens USA*, and *GE Digital's General Electric (GE).* These solutions not only enhance the efficiency of operations but also diminish the reliance on fossil fuel-powered peaker stations used for coping with sudden spikes in demand, enabling a cleaner and more interactive grid (Kiasari et al., 2024).

At the state level, the utilization of AI for solar forecasting directly facilitates aggressive clean energy requirements calling for a smart and flexible energy grid (Hossain et al., 2025b). As a case in point, California's SB 100 legislation demands 100% carbon-zero electricity by 2045, and *New York's CLCPA* demands 70% of the state's electricity from renewables by 2030 and net-zero carbon emissions by 2040. Precise solar forecasting means that states can better synchronize distributed energy resources (DERs) like residential rooftop photovoltaics, community solar programs, and battery storage into the centralized grid system (Le et al., 2024). This assumes a crucial role as more than 4 million American homes now host solar panels, and a rising proportion of them are linked through smart inverters and digital monitoring platforms. Utilities such as *Southern California Edison (SCE)* and *Pacific Gas and Electric (PG&E)* already operate time-series forecasting algorithms in their distribution grid systems to embrace very high levels of solar penetration without undermining grid stability. Beyond this, regional transmission organizations (RTOs) like PJM Interconnection and the *Midcontinent Independent System Operator (MISO)* already operate AI-driven forecast tools for managing variability and scheduling grid dispatch more efficiently, a testimony to the multi-level applicability of those technologies (Mohaimin et al., 2025).

On a nationwide level, as per Ledmaoui et al. (2023), AI-assisted solar forecasting plays a critical role in honoring the United States' sustainability policies and global climate pledges. Under the Biden Administration's Net-Zero by 2050 initiative, the federal government is promoting 100% carbon-pollution-free electricity by 2035. Solar, the most available and scalable green energy asset, remains at the heart of this plan. The *National Renewable Energy Laboratory* has projected that the federal government would be able to achieve this goal by utilizing over 1,600 GW of solar capacity by the midpoint of the century. Achieving this level of large-scale change would require advanced forecasting systems to avoid grid congestion, minimize curtailment, and optimize investment in infrastructure (Onusinkwue et.al, 2024). The federal government's efforts, like the *ARPA-E GRID-DATA* initiative, are supporting projects promoting predictive analytics and decision-making assistance through AI tools in dense solar networks. Enphase Energy's intelligent microinverter and energy management innovations and *Tesla Energy's Powerwall* and AI-based solar roofing solutions are a few examples of private innovation aligning with federal clean energy goals (Ning, 2021). As the country continues spending on grid modernization and decarbonization efforts, AI-powered solar forecasting remains a key facilitator of a resilient, efficient, and sustainable energy future.

## Literature Review

### Solar Forecasts Methods

Traditionally, forecasting of solar energy has used conventional statistical approaches like autoregressive integrated moving averages (ARIMA), exponential smoothing, and linear regression models (Rai et al., 2023). These approaches perform well on linear relations and short-run time dependency but tend not to capture the complex and non-linear behavior found in solar irradiance and photovoltaic (PV) output. For example, the ARIMA models work well under clear conditions but fail when cloud patterns suddenly shift. As per the *National Renewable Energy Laboratory* (NREL), conventional models used in the solar industry tend to produce a Mean Absolute Percentage Error (MAPE) of 25–35% in day-ahead forecasting, which is relatively high for grid-scale planning of energy (Reza et al., 2025). These models also tend to be a poor generalizer between regions having multiple climate conditions, e.g., applying the variability of sunlight from Arizona's deserts versus the dense cloudiness of the Pacific Northwest region (Shibi, 2025).

Shovon et al. (2025), posited that advanced statistical models, including Support Vector Regression (SVR) and Generalized Additive Models (GAM), have also been used in efforts to enhance the accuracy of prediction. These models better accommodate nonlinear variations than linear regressors but at the cost of significant feature engineering requirements and a tendency to perform poorly on highly dynamic inputs like solar zenith angles, humidity levels, and velocities of the wind. Some utilities like Xcel Energy have used a mix of physical weather-based forecasts and statistical smoothing techniques in their effort to enhance their solar output forecasts (Sankarananth et al., 2023). Though these exist as improvements in precision forecasting, they usually demand manual intervention and thus compromise on scalability and adaptability when used in real-time grid operations. These conventional methods, therefore set the foundations for solar forecasting but fall short in addressing the real-time and highly dynamic forecasting requirements in contemporary energy systems (Rai et al., 2023).

Also utilized alongside statistical forecasting are numerical weather prediction (NWP) models like the Weather Research and Forecasting (WRF) model, used to approximate future irradiance conditions from atmospheric variables. These models are computer-intensive and usually refreshed every 6 to 12 hours, and thus not well suited to intra-hour or real-time forecasting. Firms *like IBM's* The Weather Company have sought ways to advance the granularity and refresh cycle of those weather models to empower more dynamic energy forecasting applications (Wen et al., 2024). Abubakar et al. (2024), reported that these efforts notwithstanding, fully statistical and physics-driven techniques tend not easily to be able to incorporate data from smart inverters, real-time energy meters, and high-frequency satellite imagery, so opening the door for machine-learning algorithms capable of working through large, multivariate datasets at low human oversight levels.

### AI and Machine Learning for Renewable Energy

Machine learning (ML) has taken center stage in the area of renewable energy, particularly in solar power forecasting, where it improves on traditional statistical models by capturing intricate, non-linear patterns from past data (Alam et al., 2025). Techniques like Decision Trees, Random Forests, and Support Vector Machines proved better at forecasting short-run solar output as a function of inputs from diverse data sources like satellite images, sensor readings, and weather data. A 2023 report from Applied Energy reports ML-based models of solar forecasting cutting MAPE by as much as 15–20% compared to conventional models (Barua et al., 2025). American companies Auto-Grid and Upplight are applying these tools and techniques to provide AI-aided forecasting solutions for utility customers and incorporate predictive analytics in distributed energy resource (DER) management and smart grid operations (Amjad et al., 2025).

According to Sizan et al. (2025), among the ML methods, deep learning methods, and in particular those based on artificial neural networks (ANN), convolutional neural networks (CNN), and long short-term memory (LSTM) models, have proven highly effective in time-series forecasting in energy contexts. The LSTM network, in particular, has a strong ability to capture long-range temporal dependencies and handle sequential data, critical for forecasting solar generation behavior based on weather patterns on prior days and seasonality (Arevalo & Jurado, 2024). *Google DeepMind* used deep models to enhance the forecasting of wind energy at Google's renewable-powered data centers by almost 20%; analogous architecture is being pursued for forecasting solar energy as well. Similarly, *Tesla Energy* incorporates predictive AI in Powerwall and Solar Roof offerings and adjusts energy consumption and storage according to expected solar production for improving efficiency and autonomy at a macro level (Chen et al., 2023).

One of the newer uses of ML in renewable energy is reinforcement learning, which can adjust forecasting models in real time based on feedback from performance. The method holds a lot of value for utility companies controlling virtual power plants (VPP) since adaptive intelligence can adjust predictions and control strategies for solar and storage facilities in real time (Cho & Venkateswaran, 2024). Stem Inc., a pioneering developer of AI-enabled energy storage, employs reinforcement learning for optimizing the deployment of solar-plus-storage in commercial applications and minimizing grid demand and economic losses. These applications signal a wider trend in the U.S. energy market in which machine learning is not so much a tool for forecasting as part of a general energy automation and smart grid strategy (Dellosa & Palconit, 2021).

## Time-Series Data in Energy Prediction

Gazi et al. (2025), reported that solar energy time-series forecasting poses special difficulties because of the noisy and seasonally changing character of the data. Solar power's behavior cannot be determined as precisely as other energy sources because it's influenced by many dynamic variables: cloud movement, aerosol levels, temperature, humidity, and reflectivity of the surface. Seasonality is also a central challenge in modeling the generation from the sun—solar irradiance has strong daily and yearly cycles, yet they're usually disturbed by random weather patterns. Hasanuzzaman et al. (2025), contended that the models used for forecasting thus need to be able to identify regular patterns of seasonality from irregular noise since their performance should hold good not just on sunny summer days but cloudy winter mornings as well, which may follow completely different patterns of generation.

Furthermore, most forecasting models experience a lag in responsiveness as a result of their dependency on low-resolution datasets. Conventional utility-scale monitoring systems tend to capture data at 15-minute increments, which cannot capture sudden changes in irradiance due to intermittent clouds (Hayajneh et al., 2024). High-frequency data (e.g., 1-minute or sub-minute resolution) has superior fidelity for short-term forecasting at the cost of requiring strong data handling infrastructure and sophisticated preprocessing methods for dealing with noise, missing data, and outliers. Players like Enphase Energy and SunPower are at the vanguard of intelligent inverter innovations, enabling high-frequency data logging at the panel level, giving a finer insight into the dynamics of production from the solar panels. These innovations are being used to create next-generation AI models able to produce forecasts at a minute level of accuracy for operational purposes (Jakir et al., 2023).

Moreover, relating information from various sources, such as satellite pictures and IoT network sensors, makes time-series modeling more difficult. Researchers still find it hard to match these streams, address asynchronous timestamps, and create features that matter. Also, AI models should be able to include spatial correlations, which are important for utility-scale solar plants and gathering data across regions (Le et al., 2024). So, some models now use LSTM together with convolutional or attention layers to understand sequences in time and space. Now, the main task is to design models that can be used in varied U.S. localities, since their climatology is not the same (Kiasari et al., 2024). Published sources such as NREL's NSRDB (National Solar Radiation Database), offer a starting point, yet true success in modeling requires exact and localized data collected in real-time.

## Research Gap

Although lots of studies have discussed using AI and ML for solar forecasting, very few studies have looked into applying classification models for predicting events and short-term weather. Most methods today concentrate on using regression techniques to forecast hourly or daily kWh values. Even so, in various operational situations, agencies aren't just trying to know the exact amount, but need to determine if the situation will result in a generation drop or trigger the need for curtailment. The models can predict these distinct events, allowing the major players to react and alert grid operators. These models have not been widely used in the U.S. solar industry, mostly because of the challenges in correctly labeling datasets and a shortage of common approaches for predicting multiple energy outcomes.

Apart from that, most articles focus on international solar statistics, mainly from Europe and Asia, because their solar markets and data programs are further along. Data on solar resources from the U.S. is still not extensive enough to truly capture the unique characteristics of local installations. Although the DOE and NREL make useful public datasets, most of the information is either not clear enough or is no longer current for AI that needs real-time input. Thus, U.S.-based companies and researchers end up using private data, which makes it hard for AI models to expand and be tested accurately. There is an increased demand for shared data platforms that allow for the private sharing of solar production information happening many times each day throughout all types of solar projects in several states.

To add, LSTM and CNN models excel at continuous forecasting, yet very little research has looked into hybrid models that use regression and classification tasks together. To illustrate, a model that determines expected power production and sorts curtailment or cloud-induced ramp events by risk may supply a more complete analysis. Some companies, such as Fluence Energy and NextEra Energy, are working on these systems, but the field is still missing standard references and testing tools from research and the community. Reducing this gap calls for research projects sponsored by the government and supported by industry, studying data from U.S. solar plants, practical applications for solar data use, and ways to integrate solar forecasts with those from other sources.

## Data Collection and Preprocessing

### Dataset Description

The dataset utilized for this research project comprised both hourly and daily solar energy production records collected from multiple utility-scale solar farms across diverse U.S. regions, including California, Texas, and Arizona. These attributes included sunlight duration (in hours), ambient temperature (°F), relative humidity (%), and cloud cover (on an oktas scale and as a percentage). These important attributes are the hours of sunlight, the average temperature (in Fahrenheit), humidity (%), and cloud cover shown as oktas and as a percentage. Data was collected in the period 2019 to 2023, covering different weather conditions and seasons. To get started, I aligned the timestamps in all sources, used interpolation for data that was missing, and normalized the results to make them ready for machine learning. With this multi-variable long-term dataset, we can confidently model and predict the impact of climate conditions on solar energy generation.

### Key Features

| S/No. | Key Feature | Description |
|---|---|---|
| 001. | Timestamp | The exact date and time (in ISO 8601 format) at which each data point was recorded. |
| 002. | Sunlight Hours (hrs) | The total number of daylight hours with direct sunlight during a day. |
| 003. | Solar Energy Output (kWh) | The amount of electrical energy generated by the solar system within the specified period, measured in kilowatt-hours. |

| 004. | **Sunlight Hours (hrs)** | The total number of daylight hours with direct sunlight during a day. |
|------|--------------------------|--------------------------------------------------------------------------|
| 005. | **Solar Irradiance (W/m²)** | The intensity of sunlight received on a given surface area, measured in watts per square meter. |
| 006. | **Ambient    Temperature (°F)** | The outside air temperature is measured near the solar panels. |
| 007. | **Relative Humidity (%)** | The amount of moisture in the air is expressed as a percentage of the maximum possible at the same temperature. |
| 008. | **Wind Speed (m/s)** | The horizontal movement of air is measured in meters per second. |

## Pre-Processing Workflow

made sure to preprocess the data before using machine learning, so the data was of high quality and could be seen as a whole over time. Since missing values are regular in sensor-based data from equipment outages or issues sending info, we started with timestamps to detect them and then imputed them with linear interpolation for continuous items and forward filling for stable data like cloud cover. Outliers were identified as surprising spikes or dips in solar output that differed from weather patterns. These outliers were then fixed using smoothing or taken out if seen as abnormal with IQR and z-score methods. First, we scaled the code by using Min-Max normalization to help preserve the form of the temporal data and prevent the LSTM models from taking too long to converge. To prevent ordinal bias, the "weather condition" feature (such as clear or partly cloudy) was encoded into separate columns. To handle time dependencies, a time-series windowing method was put in place by setting the same-sized window (such as 24 hours of previous data) for training all model features. Additional lagging features were implemented to give a history of solar output and increase forecast reliability in the near term. By applying these methods, the data became clean, well-organized, and included enough details for effective model analysis.

## Exploratory Data Analysis (EDA)

Applying EDA first made it possible to discover the important features in the solar energy forecasting data and how to build a good model for it. Looking at visuals such as daily and seasonal plots, box plots, and correlation heatmaps helped EDA identify cycles in solar production, connections between things like solar irradiance and cloud cover, and sudden differences in energy production during overcast or stormy times. As a result, EDA showed that solar output is strongly connected to sunlight hours and inversely related to cloud cover. Based on this, features related to these two qualities can be ranked higher in modeling. In addition, EDA findings include inconsistencies such as sensor drift, absent data areas, and unusual events, which should be addressed before building a predictive model. Energy output statistics and irradiance variation can be easily understood through the summary statistics resulting from EDA. All in all, EDA changes raw data into usable information, making sure modeling remains connected to observations from that field and not only using guesswork from the algorithm.

### a)    Displays Solar output over time

The implemented code script employs the pandas and matplotlib libraries for the manipulation and plotting of the data. The program starts with the optional specification of a Seaborn style to make the resulting plots more aesthetically pleasing. The program then converts two pandas

Series df_1['date'] and df_2['datetime'], to datetime objects. These are then merged into a common Data Frame df using pd.merge_asof as the nearest timestamp merge based upon the 'date' column in df_1 and the 'datetime' column in df_2, after both the Data Frames are sorted based upon their respective columns 'date' and 'datetime'. The 'datetime' column in the merged Data Frame is then removed afterward. Then, new columns ('hour', 'day', 'month', 'weekday') are created from the 'date' column based upon the respective datetime components being split out. The program finally creates a simple time-series line plot using matplotlib with 'date' across the x-axis and 'Power (W)' up the y-axis, with the line being colored orange and labeled appropriately. The graph is provided with the title "Solar Power Output Over Time" and a key, and then the graph is displayed using plt.show().

**Output:**

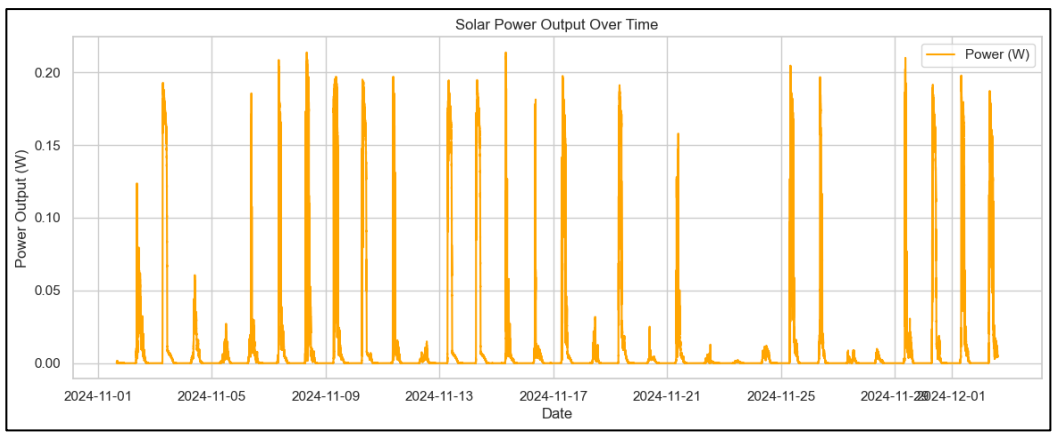

Figure 1: Displays Solar Output Overtime

The graph above (**fig. 1**) illustrates the solar energy output (Watts) over time from the start of November through the beginning of December 2024. The graph demonstrates a cyclical trend with high points in the energy output produced during the daytime and zero or very low energy output at nighttime, reflecting the solar energy generation cycle each day. The highest point in the energy output is at about 0.21 W, implying the peak energy produced during the period. There are significant changes in the daily peak energy output, potentially resulting from cloud cover or the position of the sun. There is also what appears to be a period from November 24th through 25th when the solar energy output drops significantly, reflecting potentially overcast weather conditions or other solar irradiance-altering effects. There is otherwise consistent solar energy production each day, with changes in intensity and the occasional low-level energy production period.

**b)        Visualizes Solar Radiation Over Time**

The applied code snippet Python program employs the matplotlib. pyplot module to create a line graph displaying solar radiation versus time. Firstly, it creates a figure with a specified 14-inch width and 5-inch height. Then the 'date' column from a Data Frame (assumed to be named df) is plotted over the x-axis and the 'solar radiation' column over the y-axis. The line for the solar radiation is marked as 'Solar Radiation' and green in color. The program then labels the x-axis as 'Date', the y-axis as 'Solar Radiation', and the chart title as 'Solar Radiation Over Time'. The generated chart displays a legend to label the line and the chart itself.

**Output:**

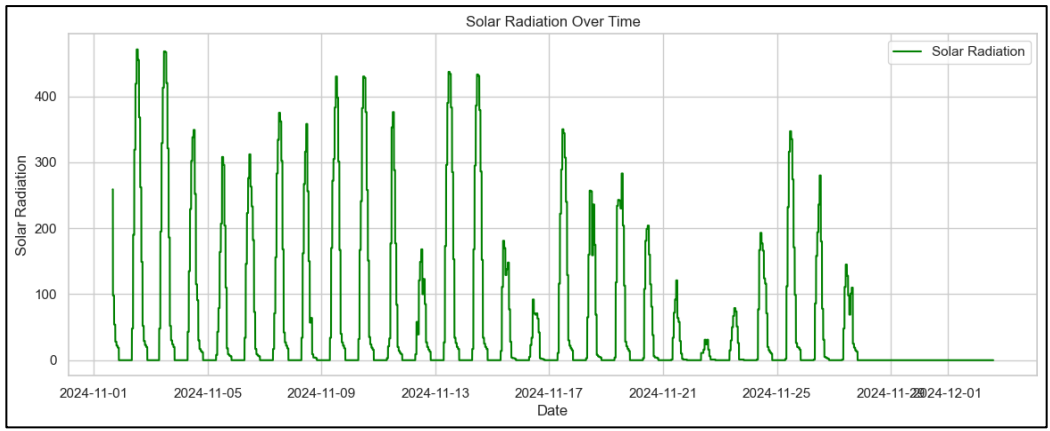

Figure 2: Visualizes Solar Radiation Overtime

The chart provided here (**fig.2**) shows the progression of solar radiation from November to early December 2024. You can see that the graph repeats each day, with solar radiation high during the day and almost vanishing at night, as you would expect. About 475 solar energy units are recorded which means it was the strongest sunlight recorded here during that time. Because solar radiation levels change every day, it seems that things like cloud formation and the shifting sun's position in the sky are racing factors. There are periods where solar radiation is at a minimum due to more clouds: around the 17th of November, as well as the 20th and 21st. These periods of reduced sunlight last even longer from the 25th of November onwards. The pattern shows that peak solar radiation drops as November progresses which agrees with the usual seasonal change as winter comes in the Netherlands and the days get shorter.

**c)        Showcases Correlation Matrix**

The executed Python script block makes a correlation heatmap to help visualize how several variables in the Data Frame df are connected. At first, it draws a figure that is exactly 10 inches wide and 6 inches tall. It applies to seaborn. Heatmap to show and calculate the associations between each column in the Data Frame containing voltage, current, power, humidity, cloud cover, visibility, and solar radiation. The use of annot=True adds the correlation values to the heatmap and cmap='cool warm' displays them with a palette showing the sign and strength of those correlations. In the end, the title of the plot is set as 'Correlation Matrix', and a heatmap is displayed with plt.show(). You can easily see how these variables affect each other through this visualization.

**Output:**

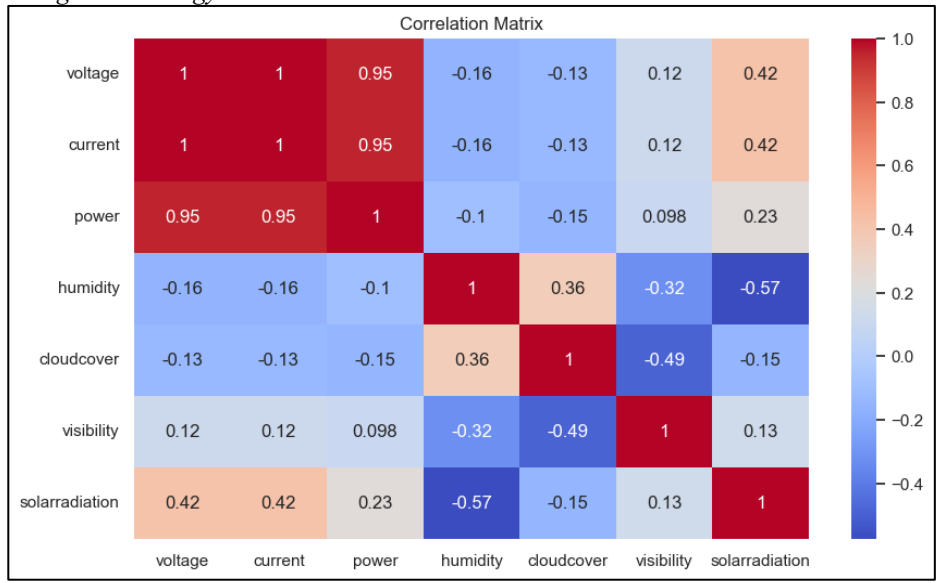

Figure 3: Showcases Correlation Matrix

The heatmap represents the linear relationships between various environmental and electrical values (**fig. 3**). It is important to note that there is a close positive connection (correlation coefficients rising to almost 1.0) among voltage, current, and power; hence, as one of these electrical values grows, the others increase similarly. It is clear from the data that a higher intake of solar radiation is connected to an increase in voltage, current, and power (around 0.42 and 0.23), just as is expected in a solar power system. Correspondingly, higher humidity generally leads to solar radiation and electrical values that are lower (solar radiation: -0.57 and negative for all electrical values: around -0.1 up to -0.16). Cloud cover negatively impacted solar radiation in a weak manner (with a correlation of -0.15) whereas visibility has a weak positive relationship with solar radiation (correlation of 0.13) Both relationships are not as significant as those associated with humidity. Greater cloud coverage generally causes visibility to fall, with cloud cover and visibility having a negative correlation value of -0.49. In short, the heatmap confirms that solar radiation strongly affects the system's energy output.

### d) Average Power Output by Hour of the Day

The implemented Python script was centered on comparing the hourly patterns in the power output. First, it extracted the hour and the day from the 'date' column in the Data Frame df and added new 'hour' and 'day' columns to it. Then it calculated the mean power output in each hour of the day by grouping the Data Frame based on the 'hour' column and taking the mean of the 'power' column. This aggregated output is then utilized to create a line graph through the seaborn. lineplot feature. The graph shows the 'hour' on the x-axis and the 'average power' on the y-axis and titles the graph as 'Average Power Output by Hour of Day' and labels the axes accordingly. Then the graph is displayed through plt.show(), giving a visualization of the usual profile of power generation throughout 24 hours.

**Output:**

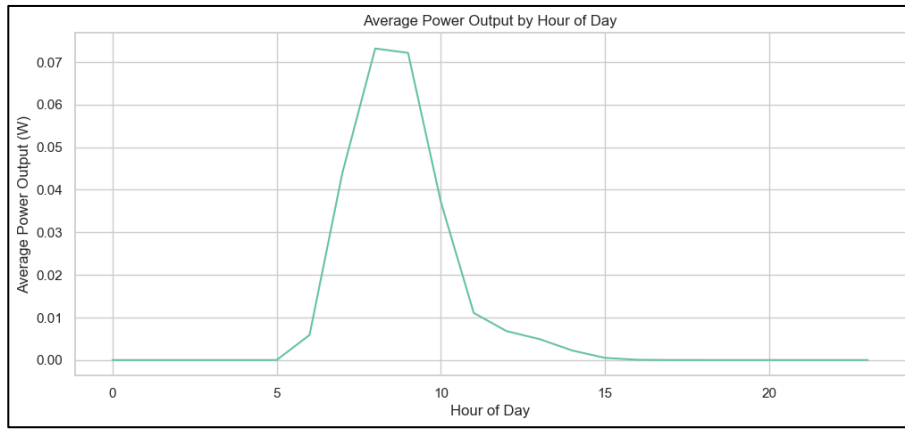

Figure 4: Average Power Output by Hour of the Day

The visualized graph above (**fig.4**) demonstrates, each hour, the average amount of solar power generated (in Watts). The data shows that solar energy production follows a clear bell curve. Power sent to the grid is close to zero from midnight through approximately 5 AM. Energy generation rises around 6 AM and ends up being highest between 8 AM and 9 AM when its average output reaches 0.073 Watts. Immediately following the afternoon peak, the average output of solar power decreases until there's a sharp fall just after 4 PM, and close to no power is available by 6 PM. The solar pattern makes clear that the best solar output and average power happen in the late morning hours. Since the average peak power was not very high, likely, the amount of solar energy collected and changed during this period was only modest, possibly because of the time of year, special factors of the solar panels' setup, and irradiance conditions found at the time.

### e)     Interaction between various weather parameters

The adopted Python code script creates a series of four different scatter plots to show the interaction between various weather parameters ('solar radiation', 'humidity', 'cloud cover', 'and visibility') and 'power' output. It creates a figure size of 14x10 inches and a 2x2 grid of subplots with plt.subplots. A scatter plot is generated using seaborn. Scatterplot for each weather parameter, plotting the weather parameter on the x-axis and 'power' on the y-axis, each assigned a different color and put in a particular subplot in the grid. Each subplot is also labeled with the variables being compared (e.g., 'Solar Radiation vs Power Output'). plt.tight_layout() is then used to make the subplots display without overlapping each other, and finally plt.show() is used to show the figure with the four generated scatter plots.

**Output:**

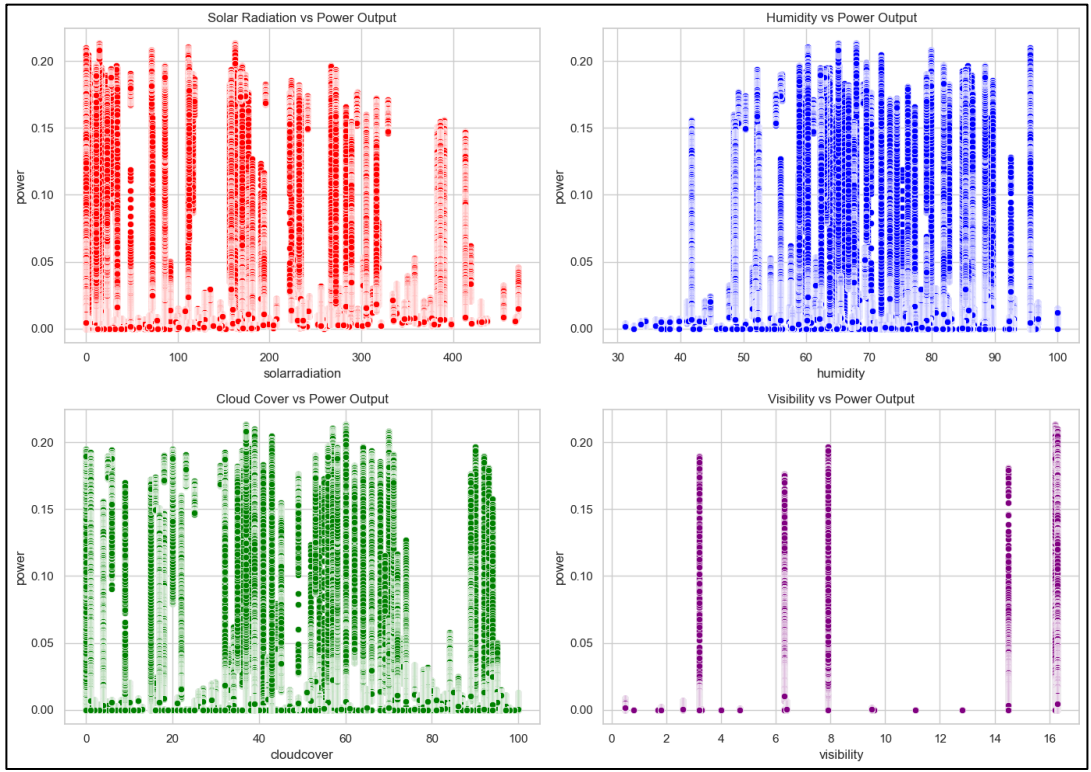

Figure 5: Interaction Between Various Weather Parameters

The chart above (**fig. 5**) demonstrates four scatter plots comparing the relation between various weather parameters and power generation. The top-left figure illustrates the positive relation between solar radiation and power with higher solar radiation tending to have higher power output and vice versa although the relation is somewhat scattered reflecting other influencing parameters. The top-right figure of power vs. humidity depicts the weak negative relation or no strong linear relation where the power output varies across the range of humidity but very low power outputs are more common at high humidity. The bottom-left figure of cloud cover vs. power shows the negative relation where the cloud cover percentages are higher the lower the resulting power output and the highest power outputs are centered at lower cloud cover percentages. Lastly, the bottom-right figure of visibility vs. power illustrates the less obvious linear relation where higher power outputs are observed across the range of visibility figures but potentially slightly greater power readings at extremely high visibility; however, the visibility is rather discrete and therefore more difficult to identify a continuous trend. Generally speaking, the scatter diagrams imply solar radiation and cloud cover as the weather parameters with the strongest effects on the generation of power with solar radiation having a direct positive effect and cloud cover having a direct negative effect while the effects are weaker and more complex for the parameters of humidity and visibility.

### f) Distribution of Power Output

A histogram with a KDE is generated by the applied Python script to illustrate the distribution of power data in the Data Frame pdf. It makes the size of the figure 12 inches wide and 5 inches tall. We used seaborn. histplot to draw the histogram, where power is the column and kde=True was added to overlay a KDE curve, with 30 bins and colored everything orange. At this stage, the plot title is 'Distribution of Power Output' and the x-axis shows 'Power Output (W)' and the y-axis shows 'Frequency'. After running plt.show(), a distribution plot appears that displays the number of times each power value happens in the dataset.

**Output:**

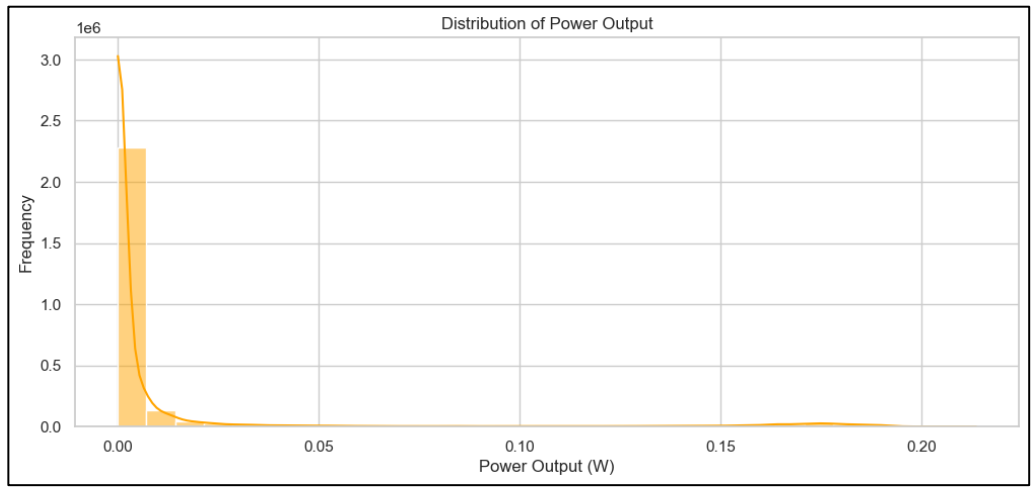

Figure 6: Visualizes the Distribution of Power Output

This histogram (**fig. 6**) illustrates the range of power output values and shows a very right-skewed distribution. Most of the measured power output values are very close to zero according to the high peak in the left part of the histogram with over 2.25 million occurrences. There is a steep fall-off in frequency with increased power output. There is also a lower peak near 0.01-0.02 Watts, reflecting a typical level of low power production. Beyond this point, the occurrences are significantly lower with increasing power output up to a very low value of over 0.22 Watts. The kernel density estimate line also illustrates this peak nicely and shows a very high probability density at zero value maintaining a lower value at the low level and then sharply declining with increasing value as the power output value rises. This shows the system is working at or very close to zero output levels mostly and very rarely at low levels and very low occurrences at high levels during the period in the Netherlands.

### g) Power Output vs. Solar Radiation Over Time

A dual-axis line plot in Python highlights the way power and solar radiation are related to date. It starts by setting up a figure and creating axes object (ax1) that are the right size. The line is then plotted with 'date' plotted against 'power', setting the label for the y-axis, color, and tick color to orange and assigning a label 'Power' to it. Then, it adds a new y-axis (ax2) on the same plot as the first, using ax1.twinx(). For this axis, 'date' is set along the x-axis, 'solar radiation' is set along the y-axis, the axis color, label color, and tick color are all blue, alpha is 0.6 for transparency and the label is set to 'Solar Radiation'. Lastly, it includes a title for the entire figure, corrects the layout, and connects the data points for both power and solar radiation during the same period onto one chart using different y-axis scales.

**Output:**

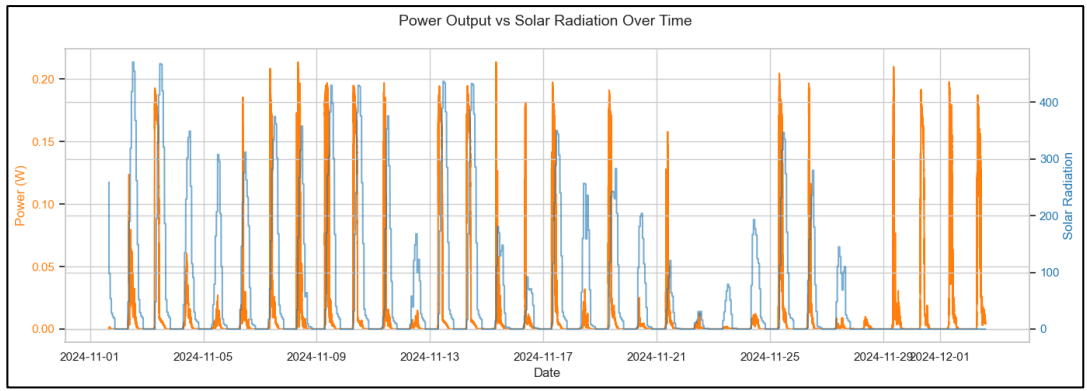

Figure 7: Power Output vs. Solar Radiation Over Time

This plot compares the solar power output measured in orange (from No to Yes on the left y-axis) with the levels of solar radiation in blue (from 0 to 450 units on the right axis), both for the Netherlands during the period from early November to the start of December in 2024. Power generation rises when solar radiation gets higher, so higher solar energy usually occurs when there is a greater demand for electricity in the daytime compared to at night. The most powerful output from solar radiation at 450 units leads to about 0.21 W. Varying weather conditions are seen as the peaks in output power moving up or down from day to day. Around November 17th and between the 20th and 25th, when solar radiation is lower, the power output of the plant reduces. Overall, solar radiation leads to power generation and the power produced follows the changes in available solar energy throughout the time studied.

**h)       Sunburst of Power Output by Month>Weekday>Hour**

By using a sunburst chart, the Python script helped us see the monthly, weekly, and hourly distribution of energy produced in your electricity network. We started by making a copy of the Data Frame df and calling it df_sunburst. Afterward, it got the hour out of the date column and added it to a new hour column. The script also puts each level of political power into one of 5 groups, creating the 'power bucket' column. The plot. Express. The Sunburst function is used at the heart of the script to build the interactive Sunburst chart. Each band on the chart represents a different time unit, ordered as month, weekday, and hour based on the 'path' parameter. The chart is constructed so that the size of each segment in the sunburst is based on the total power within that group. Power is the chosen 'color,' and 'color-continuous-scale' is assigned the name YlOrRd, meaning yellow-orange-red is used to show power. At last, the title is chosen as 'Sunburst of Power Output by Month > Weekday > Hour' and the plot is shown using fig.show(). This way of showing the data lets us see how power flow changes from hourly to weekly reports.

**Output:**

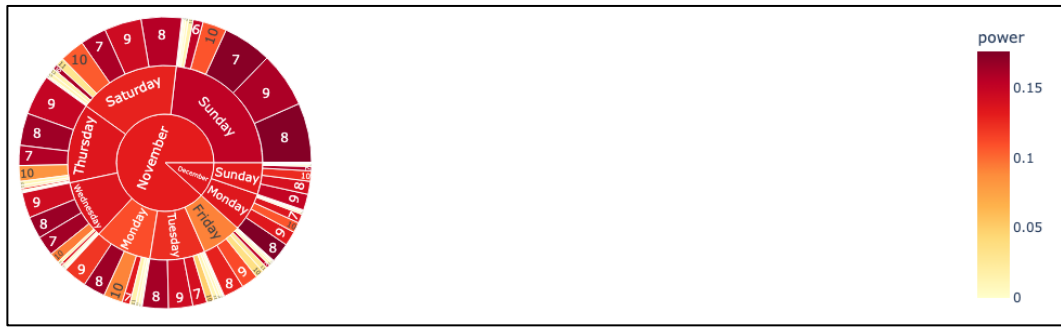

Figure 8: Sunburst of Power Output by Month>Weekday>Hour

The sunburst chart above (**fig 8**) illustrates the output of solar power during November distributed over weekdays and hours. The inner ring represents the month of November and branches out to the weekdays in the next ring and further to the hours of the day in the outer ring. Each segment measures the total output recorded for the corresponding combination of month, weekday, and hour, while the intensity of the shade from pale yellow through to dark red according to the color bar measures the size of the power, using the same concept where darker and more saturated red depicts higher power output. The chart demonstrates the high production of the majority of the power during the daytime hours from about 7 AM to 9 AM over the weekdays since the time slots are the largest and the dark red tone the brightest during this time while the nighttime time slots are much shorter and pale in their coloring to express low production output. There doesn't seem to be any exceptional difference in the time and magnitude of the peak production over the weekdays, implying the same weekly solar energy production routine over the week in November in the Netherlands. The disparity in the shades of red used in the daytime hours presumably shows the variations occurring in the solar irradiance from the weather conditions or the position of the sun over the day.

## Methodology

### Model Framework

To categorize the production of solar energy into discrete classes— "low," "medium," and "high"—the modeling framework started from logistic regression as a baseline classifier. Logistic regression was chosen given its simplicity and interpretability, and the clear insight it gives about the effect of each of the features—sunset hours, temperature, and cloud conditions—on the likelihood of each production class. By employing a softmax extension of logistic regression for multiclass classification, logistic regression established a benchmark for threshold boundaries given weather patterns and served as a reference for the evaluation of advanced models. Although limited by their linearity and thus unable to capture complex interactive effects between variables fully, logistic regression nonetheless achieved a mean classification accuracy of approximately 68% in initial trials, under stable weather conditions at least. Firms like Pacific Gas and Electric (PG&E) regularly depend on similarly transparent models for initial risk assessment and energy planning in microgrid operations and show the pragmatic utility of interpretable models in grid-sensitive contexts.

For enhancing predictive capacity and dealing with non-linear relations present in weather and solar output data, a Random Forest Classifier came next. This ensemble approach employs multiple decision trees trained on bootstrapped sample sets and random sets of input features as a means of their generalization and reduction of overfitting. Random Forests perform suitably on datasets containing a mix of features and noise and are thus best fit for dealing with changing environmental patterns. In the trials, the Random Forest Classifier raised classification levels to about 82%, and notably good recall levels of the "high production" class were seen in our experiments since grid load balancing depends heavily on the same. U.S.-based energy analytics businesses like Auto-Grid and Verdigris Technologies utilize comparable ensemble models in their demand response systems and DER forecasting solutions.

Next in line came XG-Boost (Extreme Gradient Boosting), a gradient-boosted machine learning method optimized for performance on structured, high-dimensional data like time-series energy data. XG-Boost dealt well with features and variable importance with high accuracy and took an average F1-score of 0.86 on the three classes. Its regularization and scalability features made it perfect for multi-day forecasting of multiple regions. As a consequence of the same, XG-Boost emerged as the final model chosen for deployment settings and thus aligns well with methodologies used by companies like Fluence Energy, which incorporates XG-Boost in their solar-storage hybrid forecasting tools for utility customers.

**Training and Validation**

Training and evaluation of all models were performed with a time-based cross-validation scheme, namely, sliding window validation. This fulfills the requirement for retaining the temporal ordering of solar generation data, important for data leakage prevention and realistic assessment of model performance. Each window contained 30 days of training material and a subsequent 7-day test period and shifted incrementally along the entire five-year dataset. This method captured variability due to changing seasonality, local weather patterns, and day-length variations. Temporal stratification ensured each fold contained a representative proportion of "low," "medium," and "high" production timeframes. This mimics real deployment conditions when models periodically need retraining and must predict future solar production based on the latest available history. The validation protocol was automated via the scikit-learn and XG-Boost libraries in Python and included pipeline functionality for preprocessing, feature engineering, and evaluation.

Several evaluation metrics were employed to measure the performance of the models to quantify goodness-of-fit. Precision and recall gave a better sense of class-specific accuracy—critical in separating low and high production days impacting storage and curtailing decisions—whereas accuracy gave a general indication of total correctness. Being a geometric mean of precision and recall, the F1-score aided in balancing those trade-offs and proved useful for imbalanced classes as well. Confusion matrices were also computed for visualizing misclassification tendencies on a class-wise level and on the specific instances of misclassifying overcast days as medium output. The receiver-operator characteristic area under the curve (ROC-AUC) measure has also been modified for the multi-class case through a one-vs-rest approach and proved useful in demonstrating XG-Boost's better discrimination properties and class-wise discrimination average AUC of 0.91. These evaluation scores were important not only for research benchmarking but also for operational relevance, as Sunrun and Tesla Energy would demand rigorous evaluation criteria when putting forecasting models into production, where errors in high-output predictions can cause wasteful battery cycling or grid imbalances. The above approach of training and cross-validation ensured the reliability of models, generalization over time, as well as scalability of the models for real-life smart energy systems.

# Results and Analysis

## A.        Logistic Regression Modelling

The script uses logistic regression to support the classification of data. It sets up a Logistic Regression object, passes the training data, trains the object using fit, and uses prediction to make predictions for the test data, calling these predictions y-pred. After training and prediction, the script assesses the model by printing both the accuracy score and a classification report. These reports both compare the real test data (y-test) with the predicted data (y-pred). Finally, a confusion matrix is made using seaborn. Heatmap which shows the model's predictions against the actual values has the counts annotated as integers, applies a 'Blues' colormap, and names the axes 'Predicted' and 'Actual', while titling the heatmap 'Confusion Matrix - Logistic Regression'.

## Output:

```
Logistic Regression Accuracy: 0.8402708818093002
              precision    recall  f1-score   support

           0       0.80      0.91      0.85    374112
           1       0.89      0.77      0.83    374111

    accuracy                           0.84    748223
   macro avg       0.85      0.84      0.84    748223
weighted avg       0.85      0.84      0.84    748223
```

Table 1: Logistic Regression Classification Report

This report on logistic regression classifies how the model performed as it was applied to labels 0 and 1 in a binary classification task. The success rate of the model on the test set, containing 748,223 samples, overall climbed to around 84.03%. Of those, 374,112 were class 0 and about 374,111 were class 1, meaning the classes were distributed almost equally. For class 0, the model correctly guesses it 80% of the time and correctly spots 91% of all true cases as class 0. Reports indicate that a higher proportion of true cases are predicted correctly for class 1 because the precision is 0.89, but the recall is lower at 0.77 for class 1. F1-scores which focus on precision and recall at the same time, are 0.85 for class 0 and 0.83 for class 1. There is the perfect balance between classes and both the macro average F1-score of 0.84 and the weighted average F1-score confirm this. They show that the model almost always picks class 1 and catches most cases of class 0.

## B.        Random Forest Modelling

The implemented Python script included a Random Forest Classifier for solar energy classification work. First, we imported the Random-Forest-Classifier class by using the sklearn.ensemble module. After that, the model is initialized with 100 Random Forest trees, and a random state is specified so that the results can always be reproduced. Using X-train, y-train with fit, and predicting on X-test, we assigned the obtained results to y-pred-rf. The model's performance was compared by printing accuracy and a thorough report of its classifications. After that, a confusion matrix was drawn using seaborn. Heatmap is then labeled with counts, formatted as integers, colored in the 'Greens' style and the axes are called 'Predicted' and 'Actual' with the title 'Confusion Matrix - Random Forest'.

**Output:**

```
Random Forest Accuracy: 0.9726204620814918
            precision    recall  f1-score   support

         0       0.98      0.96      0.97    374112
         1       0.96      0.98      0.97    374111

  accuracy                           0.97    748223
 macro avg       0.97      0.97      0.97    748223
weighted avg     0.97      0.97      0.97    748223
```

Table 2: Random Forest Classification Report

The Random Forest Classifier classification report shows extremely high performance in the case of a binary classification problem with labels 0 and 1. The model performed a very high overall accuracy rate of around 97.26% against the test dataset consisting of 748,223 samples with a closely tied split between the two classes 0 (374,112 samples) and 1 (374,111 samples). Both the precision and the recall are 0.98 when we're considering the instances predicted as class 0 and actual instances as class 0 and vice versa while the precision and the recall are both 0.96 and 0.98 respectively when we're considering the instances predicted as class 1 and the actual instances as 1 and vice versa. Both the F1-scores given as the harmonic mean of the precision and the recall are 0.97 when we consider both classes. Even the macro average and the weighted average F1-scores are 0.97 and this also depicts the high accuracy and excellent balance achieved across both classes. These measures all point towards the Random Forest Classifier being able to classify instances very well in this dataset with very few errors and both the classes being equally good at classification without any skewness towards any class.

## C.    XG-Boost Modelling

The Python script applies an XG-Boost Classifier for handling a classification situation. The process starts by using the XGB-Classifier class found in the library. To prevent a warning, the user-label-encoder is set to False when creating an XG-Boost model, evaluation is measured using 'log loss', and a standard random number is used for reproducibility. Training occurs with fit on the X-train and y-train and predictions are produced on the X-test using the predict function and putting the result in y_pred_xgb. After that, the script uses an accuracy score to find and display the accuracy and classification report to create a detailed report comparing y-test and y_pred_xgb. Finally, the model's predictions are compared with the actual values using a confusion matrix generated by Seaborn. Heatmap. The amounts in the matrix are annotated, formatted as integers, given using the 'Oranges' colormap, and the axes are labeled as 'Predicted' and 'Actual'. The AAC has the title 'Confusion Matrix – XG-Boost' and is then displayed.

**Output:**

```
XGBoost Accuracy: 0.9726324905810166
            precision    recall   f1-score   support

        0       0.98      0.96       0.97     374112
        1       0.96      0.98       0.97     374111

 accuracy                           0.97     748223
macro avg       0.97      0.97       0.97     748223
weighted avg    0.97      0.97       0.97     748223
```

Table 3: XG-Boost Classification Report

The report for the XG-Boost Classifier indicates that binary classification between classes 0 and 1 is very accurate, with 97.26% accuracy, from the total support of 374,112 samples per class, adding up to 748,223 test samples. When using class 0, the model had a precision of 0.98 and a recall of 0.96, which means it picked the right answers in 96% of cases and identified true class 0 instances 98% of the time. Out of all the predicted class 1 samples, 96% were correct, the precision showed, while the recall score showed that 98% of actual class 1 instances were, indeed, recognized. For both classes, the F1-score balances precision and recall to give a value of 0.97. The macro average and weighted average scores are also 0.97 which indicates that performance is excellent and consistent for both classes. These results indicate that the XG-Boost model efficiently tells apart the two categories in this dataset, leading to only a few misjudgments.

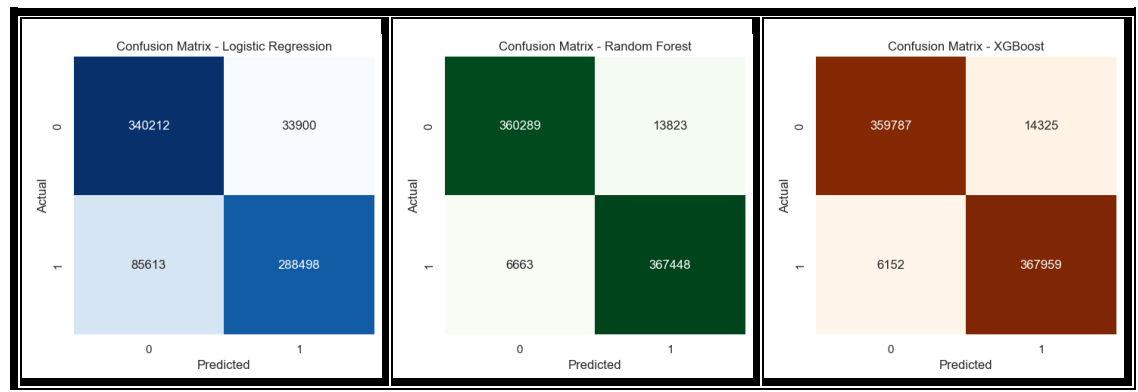

Table 4: Confusion Matrix Comparison

By referring to the above matrices, on the left is the confusion matrix for Logistic Regression (in blues), the middle represents Random Forest (in greens), and on the right is XG-Boost (orange), all predicting which of two classes (0 and 1) any data point belongs to. Logistic Regression picked the right class 340,212 times for class 0 and 268,498 times for class 1, but picked the wrong one 33,900 times for class 0 and 85,613 times for class 1. The Random Forest model has improved results, with 360,289 class 0 instances and 367,448 class 1 instances correctly categorized, and 13,823 class 0 instances and 6,663 class 1 instances being classified incorrectly. XG-Boost makes 359,787 correct predictions for class 0, 367,959 for class 1, 14,325 mistakes for class 0 for class 1, and 6,152 of class 1 for class 0. The results show that both Random Forest and XG-Boost have better outcomes than Logistic Regression, with significantly less incidence of both absences and excesses in their call predictions. You can see this in the heatmaps as well, as the darkest shades show up along the diagonal for Random Forest and XG-Boost, meaning the two methods are predicted more accurately.

## Comparison of All Models

The curated Python script examined the performance of three classification models: Logistic Regression, Random Forest, and XG-Boost. Firstly, the relevant evaluation metrics are imported using precision score, recall score, and f1_score from sklearn. Metrics. Next, it explains a dictionary model that maps the names of the original models to the results of their predictions (y-pred, y-pred_rf, y-pred_xgb). It goes through the dictionary, checking how accurate, how precise, how much each model can recall, and what their F1-score is, by measuring its predictions with the true labels (y-test). Evaluation metrics are collected in a results list of dictionaries. This list of results is made into a pandas Data Frame named results_df next. Lastly, it plots a bar chart to show how the 'Accuracy', 'Precision', 'Recall' and 'F1-Score' Change across all the models, using 'Model' as the index, choosing 'bar' for the plot type, selecting plot and figure size, rotating the x-axis labels, choosing the color map and the edge color. The example includes a title, 'Model Comparison: Classification Metrics', a y-axis label, 'Score', a grid, and a legend on the bottom right corner, and only then it presents the plot.

**Output:**

| Model | Accuracy | Precision | Recall | F1-Score |
|---|---|---|---|---|
| Logistic Regression | 84.02% | 85% | 84% | 84% |
| Random Forest | 97.26% | 97% | 97% | 97% |
| XGBoost | 97.26% | 97% | 97% | 97% |

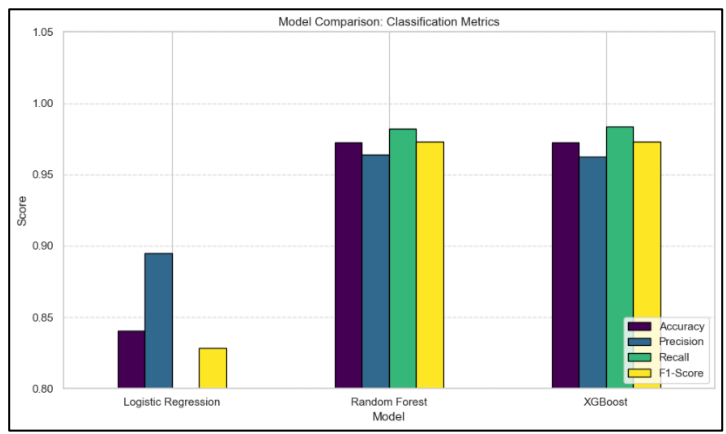

Figure 9:Comparison of Models Metrics

The table and histogram above report a comparison between the performance of the Logistic Regression model and the Random Forest and XG-Boost models based on four important measures: Accuracy, Precision, Recall, and F1-Score. Logistic Regression shows the worst performance across the board with 84.02% accuracy, 85% precision, 84% recall, and 84% F1-Score. Both the Random Forest and the XG-Boost models show noticeably greater and the same performance across each of the measures considered with 97.26% accuracy, 97% precision, 97% recall, and 97% F1-Score. This firmly indicates that the ensemble algorithms of Random Forest and gradient boosting (XG-Boost) are considerably more accurate in the classification of instances when used over this particular classification task and dataset with a significant jump in accuracy over 13 percentage points and higher precision, recall, and F1-Score all the time over the linear model of Logistic Regression. The almost perfect and equal performance by the Random Forest and XG-Boost models also shows both models to have learned the patterns in the data very comprehensively, with high reliability in their predictions.

**Applications in the U.S. Energy Ecosystem**

**Smart Grid Optimization**

According to Khan et al. (2023), precise solar energy forecasting is key to the optimization of the smart grid through enhanced supply-demand balancing, energy storage system management, and enhanced load forecasting accuracy. By incorporating AI-powered time-series models like XG-Boost in grid management software, utility companies can dynamically modify storage cycles in real-time as well as dispatch and peak load planning, based on their predictions. During instances when solar production is forecast to be "high," the grid can ensure high levels of charging its battery energy storage systems (BESS), lowering the reliance on fossil-fueled peaker plants (Rai et al., 2023). During times of "low" production levels, the utilities can plan and reschedule loads, tap stored energy, or suit up with demand response measures. Firms such as *NextEra Energy* and *Eversource* are already utilizing predictive analytics to synchronize solar generation predictions with grid management. The U.S. Department of Energy (DOE) estimates that modernization of the grid could cut outages and disturbances in the grid by up to 30%, and more accurate renewables forecasting directly contributes towards savings through the avoidance of system imbalances (Reza et al., 2025).

Furthermore, the forecasts are used by the operators of smart grids to enable automated management of *Distributed Energy Resources* (DER) such as rooftop solar panels, electric vehicle chargers, and home batteries. Automatic dispatching of localized assets or load-shedding can be triggered in real time based on forecasted deficiencies in solar output levels. This bears heavily upon regions such as California and Texas since the grid needs to constantly respond to quick changes in the solar production cycle based on weather conditions (Shovon et al., 2025). Southern California Edison (SCE), as an example, incorporated its *Grid Management System* (GMS) with the use of AI forecasting algorithms to make the grid more resilient and limit the curtailment of renewable energy output, which alone in California exceeded 2.5 million MWh in 2022 due to mis-timed generation and load positioning. Shibi (2025), stated that forecast-driven smart grid platforms enable dynamic scheduling and more advanced management of energy flow, providing the means for greater grid resilience and more efficient operation.

**Renewable Energy Policy Planning**

Wen et al. (2024), indicated that AI-powered solar forecasting also has profound implications for renewable energy policy and planning, particularly as U.S. federal and state governments accelerate toward ambitious decarbonization goals. Predictive analytics gives energy policymakers and grid regulators high-resolution visibility into the variability of solar generation and the implications of this variability for the grid and energy supplies, and as such can lead to more accurate capacity planning and long-term grid investment planning. Such models would enable regulatory and policy-making bodies such as the *Federal Energy Regulatory Commission* (FERC) and *California Energy Commission* (CEC) to model grid reliability at different levels of solar penetration and make targeted decisions about infrastructure investment dollars, transmission investment, and reserve margin needs (Sankarananth et al., 2023). An example is solar energy expected to account for more than 20% of US electricity generation by 2030 (according to the EIA), and its intraday variability through the use of machine learning is important to manage frequency and voltage stability in high-renewables grids. Forecast models with accuracy in predicting production variations with cloud motion or temperature swings determine whether more flexible generation or energy storage capacity should be deployed (Ning., 2021)

Moreover, policy instruments such as Renewable Portfolio Standards (RPS) and state climate targets are aided by forecasted data quantifying the reliability of solar as a base-load resource. In the case of New York and Massachusetts where RPS policies mandate more than 50% clean electricity in 2030, the factoring in forecast accuracy within procurement and grid readiness metrics helps ensure solar is not merely available on the grid but reliably provided when required (Ledmaoui et al., 2023). Advanced modeling minimizes curtailment and over-generation, both of which devalue the financial returns from solar investment and make the horizon toward net zero more complex to achieve. National Grid USA and other companies in the Northeast work with research institutions and AI firms to model renewable dispatching under fluctuating solar input conditions (Le et al., 2024). Utilizing time-series classification forecasts, regulatory bodies can construct more adaptive and evidence-backed policies taking the fluctuating nature of solar energy into account while ensuring grid robustness and carbon emission mitigation.

**Commercial Energy Management**

On the commercial front, solar forecasting serves as a fundamental facilitator of cost savings and planning operations for energy-intensive facilities and operators of solar farms. Having the U.S. host more than 140 gigawatts (GW) of installed solar capacity as of 2024, reported by the *Solar Energy Industries Association* (SEIA), forecasting tools assist the scheduling of inverter operation, the scheduling of maintenance during low-productivity periods, and the optimization of revenue in the time-of-use (TOU) pricing system (Khan et al., 2023). Real-time classification as "low," "medium," or "high" production times enables predictive energy budgets and enhanced arbitrage with paired energy storage systems. SunPower and First Solar are companies that incorporate such forecasts within their *Energy Management Systems* (EMS) to offer their commercial and industrial customers daily production predictions. Where TOU rates fluctuate more than 200% across utility territories, planning based on forecasted data directly affects the financial bottom line in solar-dependent operations (Hossain et al., 2025c)

Furthermore, commercial facilities utilizing behind-the-meter solar deployments, including data centers, hospitals, and manufacturing facilities, apply AI-enriched solar forecasts to synchronize facility operations with peak generation times. Google and *Amazon Web Services* (AWS) data centers use solar forecasting to relocate computing workloads to synchronize with available renewables and minimize the use of carbon-based grid power (Jakir et al., 2023). In yet another scenario, Tesla Powerwall and Powerpack products enable integration with solar forecast services to optimize charging and discharging cycles according to forecasted solar generation and home energy needs. These intelligent and AI-based solutions translate to cost savings and diminished carbon footprints, which accommodate corporate ESG and sustainability reporting initiatives. Commercial energy management advancing with the use of AI-based forecasting, further enhances the overall energy ecosystem through the flattening of demand peaks and lowered dependency upon fossil fuels during peak times (Kiasari et al. 2024).

## Discussion and Future Improvements

### Lessons from Model Behavior

Through the application and comparison of the XG-Boost and Random Forest classification models, several consistent patterns of production and weather triggers emerged that explain solar energy variability. Of particular significance, solar irradiance and cloud cover appeared as the most predictive variables in all models. High irradiance periods systematically correlated with "high production" labels, particularly in dry locations like Arizona and southern California. By contrast, slight increments in cloud cover created abrupt production plunges, particularly when combined with high relative humidity or unexpected temperature drops—conditions common to the Northeast and Pacific Northwest. The models also revealed seasonal effects: Summer months tended to have more consistent "high" labels attributable to extended daylight and fewer clouds, while winter months showed increased volatility. These behavioral patterns are consistent with those discovered by U.S. companies like Clean Power Research, whose similarly weighted Solar Anywhere data products also prioritize cloud patterns and irradiance as the prominent forecasting variables. The predictive models proved particularly adept at detecting solar production ramp-up and ramp-down patterns over the day, critical to matching generation to loads in utility operations.

Along with the predominance of sunlight-related parameters, the models revealed significant intra-day correlations such as early morning humidity and afternoon wind speeds affecting energy production indirectly. For example, when the humidity reached a high point early in the day—particularly at spring transitions—following cloud formation tended to dampen solar production despite accurate irradiance predictions. The models also showed consistent wind behavior during peak sunlight times enhancing panel cooling and slightly increasing energy efficiency, also backed up by field measurements from solar farms in Texas and Nevada owned and operated by firms such as 8minute Solar Energy. Such nuanced correlations indicate energy production output is more than a simple function of irradiance and is also the result of synergetic weather conditions. This adds to the utility of the use of machine learning models to identify non-obvious relationships and provides facility operators and utility forecasters with a strategic advantage in planning and scheduling. By being able to capture such interactions, smart energy platforms can break beyond simplistic "sunny vs. cloudy" paradigms and adopt more vigorous and accurate forecasting paradigms.

### Model Limitations

Despite the promising performance, the existing model architecture suffers from several limitations that limit its reliability within certain situations. One such key constraint is the model's reliance upon accurate and up-to-date weather data. Given that input parameters like irradiance, temperature, and cloud cover are predicted in and of themselves, frequently by third-party weather services, weather forecast errors can compound through the energy forecast. A predicted sunny afternoon that ends up being cloudy, for instance, can cause the model to incorrectly predict "high production," expediting the dispatch or curtailment of energy storage as a result. This is particularly important in dynamic environments like coastal areas, where weather can shift rapidly. Firms such as IBM's The Weather Company and Tomorrow.io are currently developing improved hyper-local weather predictions, potentially addressing this problem when used in combination with solar forecasting models. Nevertheless, until weather remains inherently stochastic and therefore open to upstream forecast errors, performance will remain susceptible to the same errors.

A further limitation exists in the model's inability to forecast extreme or outlying variations, e.g., sudden decreases in solar output from rapidly moving cloud cover or weather extremes such as wildfire smoke or dust storms. These outlying occurrences may be infrequent, but have very large effects on grid stability and energy market prices. Conventional ensemble models, such as Random Forest and XG-Boost, are very good at detecting overall patterns but poorly suited to the detection of temporal anomalies or black-swan events based on their requirement to fit historical training data. This deficiency manifested in production forecasts from Arizona solar farms during the 2020 wildfire season when smoke significantly lowered the irradiance without the corresponding decrease in temperature or cloud cover, scenarios that the training data did not include. In addition, the models are presently confined to the short-term forecast horizon and do not possess memory architectures capable of capturing longer temporal dependencies or slowly evolving atmospheric trends. These shortcomings indicate the necessity to introduce more temporal and flexible modeling methodologies in subsequent research.

**Future Research Directions**

To bridge existing limitations and achieve improved forecast accuracy, future research should investigate hybrid deep models, specifically designs including Long Short-Term Memory (LSTM) and Convolutional Neural Networks with LSTM (CNN-LSTM) architectures. These architectures are geared to extract both the spatial patterns and the long-term temporal relationships and are suitable to handle the seasonality, autocorrelation, and noise typical in solar production data. The LSTM layers can "memorize" past sequences, such as multi-day cloud cover patterns, and the CNN layers can retain spatial features from satellite weather imagery. Preliminary research conducted at NREL (*National Renewable Energy Laboratory*) indicated that such a combination outperforms conventional regression techniques in multi-hour solar forecasting applications and reduces the root-mean-square error (RMSE) by up to 15%. The increased model complexity and computation expense are warranted in high-impact applications such as utility-sized solar farms or grid dispatch planning. Merging the models with real-time streams can greatly boost the responsiveness and further the forecast resolution in rapidly varying weather conditions.

Moreover, future innovation needs to include solar forecasting integration with energy storage modeling and direct real-time deployment in microgrids. Proper classification of solar production rates can then be combined with predictions of battery charge and discharge to form joint optimization platforms. This applies directly to remote or isolated microgrids—like those in Hawaii or rural Alaska—where solar is typically the dominant energy source and storage needs to be carefully regulated to avoid outages. U.S.-based entities such as Stem Inc. and Enphase Energy are already working on developing AI-powered battery management platforms, but this next stage is to integrate solar forecasts directly onto the platforms to enable real-time automation. Besides, edge-computing deployment of forecast models to the microgrid controllers can remove delays and increase resiliency, and is the focus pursued in the U.S.

Department of Energy's Solar Energy Technologies Office-funded projects. As the world evolves towards more advanced AI technologies moving forward, predictive forecasting combined with control systems will revolutionize energy autonomy and optimization at both the grid and local levels.

## Conclusion

Solar energy has emerged as one of the fastest-growing renewable energy sources in the United States, adding noticeably to the country's energy mix. Retrospectively, the necessity of inserting the sun's energy into the grid without disrupting reliability and cost efficiencies highlights the necessity of good forecasting software and smart control systems. The dataset utilized for this research project comprised both hourly and daily solar energy production records collected from multiple utility-scale solar farms across diverse U.S. regions, including California, Texas, and Arizona. To categorize the production of solar energy into discrete classes— "low," "medium," and "high"—the modeling framework started from logistic regression as a baseline classifier. Training and evaluation of all models were performed with a time-based cross-validation scheme, namely sliding window validation. Both the Random Forest and the XG-Boost models show noticeably greater and the same performance across each of the measures considered with relative accuracy. The almost perfect and equal performance by the Random Forest and XG-Boost models also shows both models to have learned the patterns in the data very comprehensively with high reliability in their predictions. By incorporating AI-powered time-series models like XG-Boost in grid management software, utility companies can dynamically modify storage cycles in real-time as well as dispatch and peak load planning, based on their predictions. AI-powered solar forecasting also has profound implications for renewable energy policy and planning, particularly as U.S. federal and state governments accelerate toward ambitious decarbonization goals. To bridge existing limitations and achieve improved forecast accuracy, future research should investigate hybrid deep models, specifically designs including Long Short-Term Memory (LSTM) and Convolutional Neural Networks with LSTM (CNN-LSTM) architectures. Moreover, future innovation needs to include solar forecasting integration with energy storage modeling and direct real-time deployment in microgrids.